\documentclass[%
reprint,
groupedaddress,
amsmath,amssymb,
superscriptaddress
]{revtex4-1}
\usepackage[english]{babel} 
\usepackage{graphicx} 
\usepackage{dcolumn}
\usepackage{bm}
\usepackage{amsmath}  
\usepackage{amssymb} 
\usepackage{booktabs} 
\usepackage{adjustbox} 
\usepackage{color, colortbl} 
\usepackage{diagbox}
\usepackage{svg}
\usepackage[normalem]{ulem}
\usepackage[a4paper,left=3cm,right=3cm,top=2.5cm,bottom=2.5cm]{geometry}

\makeatletter
\newcommand{\mytag}[2]{%
	\text{#1}%
	\@bsphack
	\protected@write\@auxout{}%
		{\string\newlabel{#2}{{#1}{\thepage}}}%
	\@esphack	
}
\makeatother

\usepackage{xcolor}   
\colorlet{RED}{red}

   \makeatletter
     \renewcommand\@make@capt@title[2]{%
      \@ifx@empty\float@link{\@firstofone}{\expandafter\href\expandafter{\float@link}}%
       {\textbf{#1}}\@caption@fignum@sep#2\quad}%
     \makeatother
 
\makeatletter 
\renewcommand{\fnum@figure}{\textbf{Figure~\thefigure}}
\makeatother

\begin{document}
	
\title{Extraction of the propulsive speed of catalytic nano- and micro-motors under different motion dynamics }
\author{Rafael Mestre}
\thanks{These two authors contributed equally}
\affiliation{Institute for Bioengineering of Catalonia (IBEC), The Barcelona Institute of Science and Technology (BIST), Baldiri I Reixac 10-12, 08028 Barcelona, Spain.}
\author{Lucas S. Palacios}
\thanks{These two authors contributed equally}
\affiliation{Institute for Bioengineering of Catalonia (IBEC), The Barcelona Institute of Science and Technology (BIST), Baldiri I Reixac 10-12, 08028 Barcelona, Spain.}
\author{Albert Miguel-L\'{o}pez}
\affiliation{Institute for Bioengineering of Catalonia (IBEC), The Barcelona Institute of Science and Technology (BIST), Baldiri I Reixac 10-12, 08028 Barcelona, Spain.}
\author{Xavier Arqu\'{e}}
\affiliation{Institute for Bioengineering of Catalonia (IBEC), The Barcelona Institute of Science and Technology (BIST), Baldiri I Reixac 10-12, 08028 Barcelona, Spain.}
\author{Ignacio Pagonabarraga}
\affiliation{Departament de Física de la Matèria Condensada, Universitat de Barcelona, 08028 Barcelona, Spain.}
\affiliation{Universitat de Barcelona Institute of Complex Systems (UBICS), Universitat de Barcelona, 08028 Barcelona, Spain.}
\affiliation{CECAM, Centre Européen de Calcul Atomique et Moléculaire, École Polytechnique Fédérale de Lausanne, Batochime, Avenue Forel 2, 1015 Lausanne, Switzerland.}
\author{Samuel S\'{a}nchez}
\email{ssanchez@ibecbarcelona.eu}
\affiliation{Institute for Bioengineering of Catalonia (IBEC), The Barcelona Institute of Science and Technology (BIST), Baldiri I Reixac 10-12, 08028 Barcelona, Spain.}
\affiliation{Institució Catalana de Recerca i Estudis Avançats (ICREA), Pg. Lluís Companys 23, 08010 Barcelona, Spain.}
\date{\today}

\begin{abstract}

Motion of active particles, such as catalytic micro- and nano-motors, is usually characterized  via either dynamic light scattering or optical microscopy. In both cases, speed of particles is obtained from the calculus of the mean square displacement (MSD) and typically, the theoretical formula of the MSD is derived from the motion equations of an active Brownian particle. One of the most commonly reported parameters is the speed of the particle, usually attributed to its propulsion, and is widely used to compare the motion efficiency of catalytic motors. However, it is common to find different methods to compute this parameter, which are not equivalent approximations and do not possess the same physical meaning. Here, we review the standard methods of speed analysis and focus on the errors that arise when analyzing the MSD of self-propelled particles. We analyze the errors from the computation of the instantaneous speed, as well as the propulsive speed and diffusion coefficient through fittings to parabolic equations, and we propose a revised formula for the motion analysis of catalytic particles moving with constant speed that can improve the accuracy and the amount of information obtained from the MSD. Moreover, we emphasize the importance of spotting the presence of different motion dynamics, such as particles with active angular speed or that move under the presence of drift, and how the breaking of ergodicity can completely change the analysis by considering particles with an exponentially decaying speed. In all cases, real data from enzymatically propelled micro-motors and simulations are used to back up the theories. Finally, we propose several analytical approaches and analyze limiting cases that will help to deal with these scenarios while still obtaining accurate results.
\end{abstract}

\maketitle

\onecolumngrid
\section{Introduction}

The nature of living organisms is to be constantly out of equilibrium, transforming chemical energy into movement or force in order to survive. There has been extensive research to understand the physical behaviour of these organisms, from the macro- \cite{Sharpe2015,Cox2016,Brown2018}, to the nano-scale \cite{Baskaran2009,Persat2015,Aznar2018,Chen2018}. Usually, the complexity of these organisms makes difficult the study of the fundamental aspects of their dynamics, particularly in micro- and nano-organisms, whose interest has increasingly grown over the years. Recent advances in material science have led to the creation of bio-mimetic synthetic particles \cite{Hanggi2009,Buttinoni2012,Sanchez2015,Tu2016,Mano2005,Pantarotto2008}, also called self-propelled particles since they can actively create their own motion. These systems reduce the level of complexity of biological organisms, while still allowing us to study the fundamentals of motion at the nano- and microscale, which is the basis for understanding and controlling their different functionalities. Because of the large interest drawn by the active particle field, in the latest years many articles have appeared claiming the fabrication of new kinds of active particles following different strategies, usually called micro- or nano-motors. For example, some researchers use inorganic catalysts by means of metallic elements to propel the particles \cite{Katuri2016}, while others use organic catalysts by exploiting enzymes to power their propulsion \cite{Ma2015,Patino2018,Arque2019,Dey2015a,Abdelmohsen2016a,Joseph2017,Dey2015a,Schattling2015,Nijemeisland2016c}. The emergence of a wide variety of sizes and shapes has also increased the richness of the field, expanding the range of possible fundamental studies. For example, many authors use spherical, porous \cite{Hortelao2017a, Hortelao2019}, solid \cite{Patino2018a} or hollow \cite{Arque2019,Nijemeisland2016c} particles, while others use cylindrical motors to mimic bacteria \cite{Vilela2017}, star-like structures to mimic viruses \cite{Kim2018} or much more complex structures \cite{Tu2016,Wang2019a}.\\

Although there is a common path or convention for the motion analysis of micro- and nano-motors \cite{Patino2018a,Howse2007a,Novotny2019}, there could be concerns regarding the validity of the approximations or methods, the accuracy of the results and  what to do when there are slight deviations from the initial assumptions of the theory, particularly when ergodicity is broken and the micro- or nano-motor does not behave as a classic active Brownian particle. Previous works have dealt with the statistics arising from these kinds of systems \cite{Babel2014,Sevilla2014}, paying attention to the presence of convection flows \cite{Dunderdale2012a,Byun2017}, their tracking \cite{Saxton1997,Saxton1994}, the reproducibility of the experiments \cite{Novotny2019a} and their analysis strategies \cite{Kepten2015a,Gal2013}, among many others. However, there is still a lack of consistency in the literature when reporting parameters like the speed of the motors, relying on different computational methods that might represent different physical scenarios and might not be comparable with each other. Considering this, here we focus on the different ways of reporting this metric, for instance in the form of instantaneous speed or propulsive speed from MSD fittings. We compare the validity of these approximations to the speed of the particle,  evaluate the limits and sources of errors of such approximations and  provide an alternative analysis that could potentially decrease the magnitude of the error and increase the amount of information extracted. Then, building from that, we expand this reasoning towards more complex cases related to the calculation of the speed, such as when ergodicity is broken by a non-constant velocity profile, when there is active rotational speed or when there are drift currents in the system. By analyzing these cases, we provide several approaches and approximations to tackle the problem of complex speed dynamics, in order to obtain more reliable outcomes to be able to compare results from different publications and work towards a standardized common path for the analysis of motion of micro- and nano-motors.

\section{Materials and methods}

The simulations were carried out in Python assuming the description of an Active Brownian Particle (ABP) in two dimensions, as reported elsewhere \cite{Volpe2014b}. The ABP model follows:

\begin{equation}\label{eq:ABP} 
\begin{aligned}
\dot{x}(t) &= v_p\cos(\theta(t)) + \sqrt{2D_t}\xi_x(t), \\ 
\dot{y}(t) &= v_p\sin(\theta(t)) + \sqrt{2D_t}\xi_y(t), \\ 
\dot{\theta}(t) &= \sqrt{2D_R}\xi_\theta(t), \\ 
\end{aligned}
\end{equation} \\

\noindent
where $x$, $y$ are the Cartesian coordinates and $\theta$ the polar coordinate, $\dot{x}$, $\dot{y}$, $\dot{\theta}$ are their time $t$ derivatives, $D_t$ is the translational diffusion coefficient, $D_R$ is the rotational diffusion coefficient, $v_p$ is the propulsive speed of the particle and $\xi_i(t)$ is an uncorrelated Gaussian white noise with $\left\langle\xi_i(t)   \right\rangle = 0$ and   $\left\langle\xi_i(t)\xi_j(s)   \right\rangle = \delta_{ij}\delta(t-s)$. In order to simulate the most typical experimental conditions, the simulation time step was set to $\Delta t = 0.04$ ms, that is, 25 frames per second (FPS). A simulated experiment consisted on 30 particles simulated for 30 s, from whose trajectories the mean square displacement (MSD) was computed. To compute the errors of the Taylor approximation fittings, 100 experiments were simulated, that is, 100 sets of 30 particles for 30 s, in order to estimate how reproducible is the deviation of the experimental parameters from the theoretical value in different sets of experiments. \\

Real experiments of micro-motors where performed with urease-coated hollow silica micro-particles. Briefly, the silica micro-particles were synthesized from 2 $\mu$m polystyrene particles by mixing them with an ammonium hydroxide solution, 3-aminopropyltriethoxysilane (APTES) and tetraethylorthosilicate (TEOS) in ethanol solution. The polystyrene beads with a silica shell were washed with dimethylformamide to remove the polystyrene core, making them hollow. Functionalization of the micro-motors with urease was performed by functionalizing first with gluteraldehyde followed by urease from \textit{Canavalia ensiformis} (Jack bean) \cite{Arque2019}. The videos of the enzymatic micro-motor motion were recorded using the camera (Hamamatsu Digital Camera C11440) of an inverted optical microscope (Leica DMi8). A 63x water immersion objective (NA = 1.2) was used to record the micro-motors at 25 FPS, placed on a glass slide, mixed with urea at a final concentration of 100 mM (Sigma) and covered with a cover slip. For the long time tracking of the micro-motors, videos of 30 seconds were recorded at the beginning of each minute from 0 up to 10 minutes, being the minute 0 right after mixing the micro-motors with the substrate solution. From second 30 to 60, the recording stopped and that time was used to refocus the particle in the field of view. In the case of drifting particles, the drift was generated by making the solution touch the lateral part of an adhesive tape attached the glass slide, creating a slow flow in the solution.

\section{Results and discussion}

\subsection{Calculation of the instantaneous speed from particle trajectories}\label{sec:Errors}

Motion of micro- and nano-motors can be described by their translational diffusivity, $D_t$, their rotational diffusivity, $D_R$ (or its inverse, the rotational diffusion time, $\tau_r$), which are controlled by the random fluctuations of the medium\cite{Chandrasekhar1943}, and their propulsive speed, $v_p$, if there is a source of activity (like catalytic or light-driven) to self-propel the particle. The mean square displacement (MSD) of the particle's trajectory (from optical tracking experiments) is the most used method to extract this parameters to characterize the motion of micro- and nano-motors, following the well-known formula:

\begin{equation}
		\textrm{MSD}( t) = 4D_t  t  + 2v_p^2\tau_r^2\left(\frac{ t}{\tau_r}+e^{-\frac{ t}{\tau_r}}-1\right),\label{eq:MSDactive}
\end{equation}

\noindent where $ t$ is the elapsed time \cite{Howse2007a}. This result comes from performing ensemble and time averages on sets of trajectories (see Supplementary Information for more details) and it assumes that all particle trajectories are equivalent, that is, the propulsive speed of the particle is constant in time. By performing a fitting to this equation, or to simplified versions that will be considered later on, we can extract the value of the parameters to characterize the motion.\\

\begin{figure}[!t]
        \center{\includegraphics[width=0.95\textwidth]{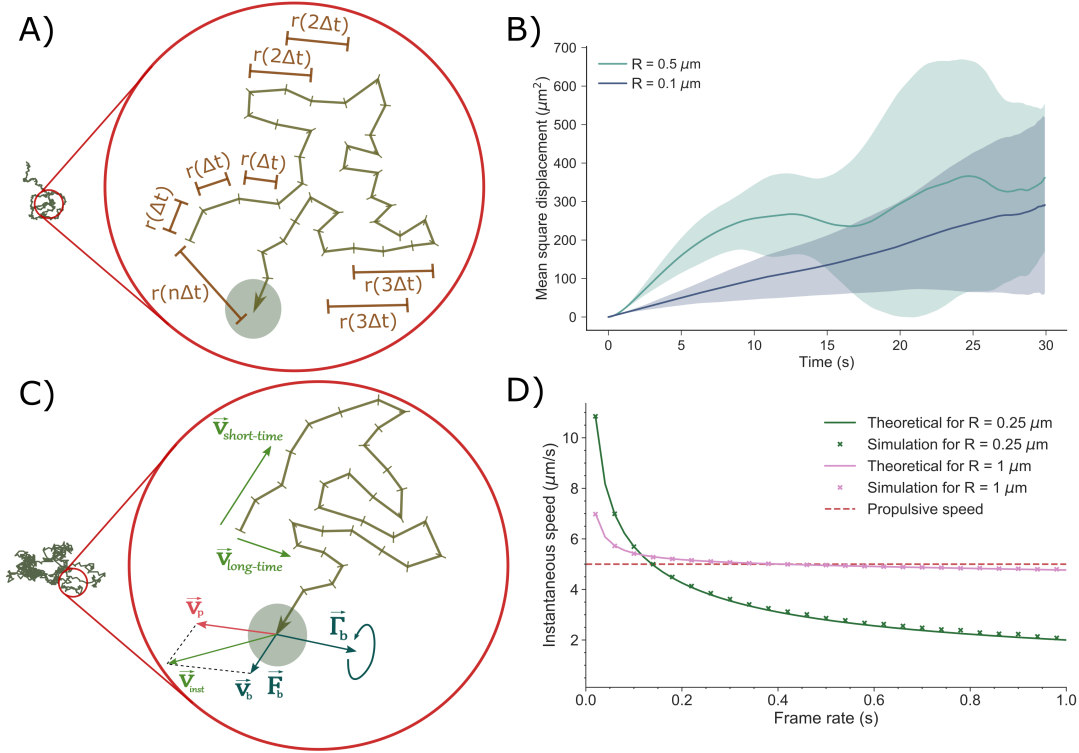}}
        \caption{\label{fig1}Considerations on the variability and instantaneous speed of self-propelled particles. A) Schematic of the calculation of the time-averaged MSD (TAMSD) of a particle trajectory. B) Average MSD of 30 simulated active particles with trajectories of 30 s of different sizes with a speed of $v = 5\, \mu$m . Shaded regions correspond to the standard deviation of the average MSD. C) Schematic of the calculation of the instantaneous speed of a particle. D) Instantaneous speed vs. frame rate for two particles of two different sizes.}
\end{figure}

From the perspective of the MSD, both ensemble and time averages are helpful to reduce the error associated to the inherent stochasticity of the Brownian process. However, because the trajectories of the particles are finite, the error associated to the MSD calculation increases with the length of the time step considered. During the calculation, the longer the time displacement, the less data points are available in the whole trajectory. We can illustrate this with the example from Fig. \ref{fig1}A. Experimentally, by using optical microscopy to measure the position of the particles, the smallest time increment, $\Delta t$, that we can obtain is equal to the inverse of the FPS. In this way, a series of points, each one taken after a time $\Delta t$, define the particle trajectory.  Many small displacements $\textbf{r}(\Delta t)$ can be obtained from this trajectory, but the displacement  $\textbf{r}(n\Delta t)$ will be only one point. Therefore, the calculation of the MSD has great inherent variability that increases with time due to the decreased number of data points available. \\

In order to test the variability of the MSD, we simulated a typical experimental case of 30 trajectories of 30 s following the ABP model (Eq. \ref{eq:ABP}) with $\Delta t = 40$ ms, or 25 FPS, which could be considered a typical experimental design in the field \cite{Patino2018a,Arque2019,Patino2018,Ma2015}. We chose the ABP model since it has been proven to be a good approximation for the motion of an active particle in 2 dimensions \cite{Volpe2014b}. In Fig. \ref{fig1}B we can see that the standard deviation of the MSD gets wider with increasing time and the mean value cannot be fully trusted for high $\Delta t$ due to the low statistical power. In general, the convention is to consider steps up to $T/10$, where $T$ is the total length of the video, where the variability is low, and to average as many particles as possible performing also an ensemble average, if the system is ergodic. The total number of particles will strongly depend on the system and the systematic errors from recording and tracking of the particles, therefore the standard deviation of the average MSD should be calculated in order to decide whether the amount of error at a specific time-point is acceptable. \\

Normally, the propulsion speed, $v_p$, is presented as the main parameter to characterize the motion of micro-motors. However, its calculation can differ from study to study. One of these strategies is calculating the instantaneous speed of the particle at each time point by performing a finite differentiation of the position of the particle with the central finite difference method of first order, defined as:

\begin{equation}\label{eq:instSpeedFiniteDiff}
\textbf{v}(t,\Delta t)=\frac{\textbf{r}(t-\Delta t)-\textbf{r}_i(t+\Delta t)}{2\Delta t},
\end{equation}

\noindent
where $\textbf{v}$ is called the instantaneous velocity of a particle. Unfortunately, using this formula as an estimation of the propulsion speed can result in under- or over-estimation of the value and should only be used in very specific situations, where motion is very directional, but always being aware of and acknowledging the limitations \cite{Palacios2019,Xuan2018,Katuri2018,Katuri2018a}. The reason to avoid this method lays on the sources of motion that affect the dynamics of an active particle (Fig. \ref{fig1}C). On the one hand, the random fluctuations due to Brownian motion  produce a rotation on the particle (which we illustrate by a torque $\Gamma_b$) and a Brownian force ($F_b$) that will have an associated Brownian speed ($v_b$). This Brownian speed can only be appreciated at times much smaller than the momentum relaxation time of a particle, $\tau_p$, where the motion is still dominated by inertia and is ballistic \cite{Uhlenbeck1930}. Therefore, it can only be measured at very high frequencies, with very complex optical setups, and not by standard optical microscopy, the temporal resolution of which is, most of the times, not high enough \cite{Li2010,Brites2016}. On the other hand, a catalytic nano- or micro-motor possesses a propulsion speed, $v_p$, independently of the underlying mechanism of motion \cite{Illien2017,Golestanian2005b,Golestanian2007b}. Hence, in optical microscopy, unless measuring at very high frequencies, the intrinsic self-propulsion will be always entangled with the Brownian diffusion, and the results obtained with Eq. \ref{eq:instSpeedFiniteDiff} will represent a convolution of both mechanisms (Fig. \ref{fig1}C). That calculation of the instantaneous speed has an intrinsic limitation, since at low $\Delta t$ the translational diffusion can dominate the magnitude of the speed, except for $D_t \rightarrow 0$. Indeed, we can prove that, if we define the modulus of the instantaneous speed as $\hat{v}_{inst} = \sqrt{v^2_x + v^2_y}$ and we average over time (assuming it  is constant in the ABP model), we obtain the following average instantaneous speed:

\begin{equation}\label{eq:instSpeed}
\begin{aligned}
<\hat{v}_{inst}^2(\Delta t)> = \frac{<r^2(2\Delta t)>}{4\Delta t^2}>=\frac{2D_t}{\Delta t} + \frac{1}{2}\left(\frac{v_p\tau_r}{\Delta t}\right)^2 \left(\frac{2\Delta t}{\tau_r} + e^{-\frac{2\Delta t}{\tau_r}} - 1\right).\\
\end{aligned}
\end{equation}

This equation shows that the instantaneous speed of the particle is not equal to the propulsive speed, $v_p$, and, in fact, it depends on the frame rate of the video, which is in the inverse of the frames per second ($\Delta t = 1/\mathrm{FPS}$). Therefore, the reported instantaneous speed is a value that strongly depends on the characteristics of the optical microscope and is not a reliable metric to compare results from different publications. Fig. \ref{fig1}D shows that explicit dependency on the time frame. We can see, for instance, how increasing the FPS of a video (decreasing the $\Delta t$) to get better temporal resolution, would only make the estimation of the speed diverge from its real propulsive speed, indicated by the red dashed line. Likewise, low recording FPS would create and underestimation of the value. These effects are much stronger when we deal with a nano-motor instead of a micro-motor, due to the more intense Brownian fluctuations of the former. Instantaneous speed of micro-motors, which follow more directional trajectories under the ABP theory\cite{Howse2007a}, can be more reliable if the FPS is not too small. In any case, the real $v_p$ could be theoretically calculated by performing a fitting to the previous equation,  this method can be more complex and would require previous knowledge of the system parameters, not being very transparent. For these reasons, calculating the speed of the particle through the instantaneous speed can be error-prone and might not give useful estimations if the system departs from the ABP model. Moreover, systematic errors arising from the recording and tracking of the particle will make this effect even more relevant, since small deviations in the particle position for slow moving particles will overestimate the value of its speed. \\

\subsection{Calculation of the propulsion speed from MSD fittings}\label{sec:constant}

The second method to calculate the speed of an active particle is through an MSD fitting to extract the propulsive speed. This is the most robust method, given that the MSD is calculated over the trajectories of many particles and time-averaged to reduce its variability. To obtain this parameter, together with $D_t$, the MSD is fitted to Eq. \ref{eq:MSDactive}. However, the motion described by Eq. \ref{eq:MSDactive} can be separated into propulsive and enhanced diffusion regimes, which simplify the fitting procedure. At short times compared with the rotational diffusion time ($ t \ll \tau_r$), particles  behave ballistically and the MSD is parabolic. Howse \emph{et al.} \cite{Howse2007a} proposed a Taylor expansion to obtain a simple approximation to the MSD for active particles in their propulsive or ballistic regime, obtaining $\mathrm{MSD}( t)=4D_t  t+v^2  t^2$. At longer times ($ t\gg \tau_r$), the behavior of particles is diffusive, and hence, taking the limit $t\gg \tau_r$ in Eq. \ref{eq:MSDactive} shows that $\mathrm{MSD}(t)=4D_e t$, where $D_e$ is called the enhanced diffusion. This value, in theory, will be larger than $D_t$ due to the effect of the activity of the particle and reflects that active Brownian particles, at longer time scales, cover a wider area than passive ones. \\

Splitting Eq. \ref{eq:MSDactive} into two regimes is always a valid approximation, independently of the size of the particle. However, the threshold defined by $\tau_r$ depends on the radius as $\tau_r = \frac{8\pi \eta r^3}{k_BT}$, where $\eta$ is the viscosity of the medium, $k_B$ is the Boltzmann constant and $T$ is the temperature. If the particle is small, $\tau_r$ will be close to zero, and the temporal resolution of a microscope will not make it possible to observe the ballistic regime. This will be the case for particles approximately below $R=0.5\, \mu$m, where $\tau_r = 0.69$ s for room temperature conditions in water ($T = 293$ K).  In these cases, only the diffusion, and not the speed, can be used to characterize the motion of the self-propelled particles, since the latter cannot be extracted from the MSD and the instantaneous speed does not represent any real value. For larger particles, of $R = 3\, \mu$m or more, $\tau_r$ can be higher than the typical observable times ($\tau_r = 149.29 $ s) and the enhanced diffusive regime might be too far in time to be visible. However, for particles of around $R = [0.5, 1]\, \mu$m of radius, $\tau_r$ can be within the observable regime, which is in the order of seconds. In this case, choosing one of the two equations might not be possible, as the approximation limits ($t \ll \tau_r$ and $t\gg \tau_r$) might not be valid anymore. In order to circumvent this problem, we found that expanding the MSD approximation to higher polynomial orders when doing the Taylor expansion leads to more interpretable results. Thus, if we expand the  MSD of Eq. \ref{eq:MSDactive} up to fourth order:

\begin{equation}\label{eq:taylor}
\textrm{MSD}(t)=\underbrace{\underbrace{\underbrace{4D_t t+v_p^2 t^2}_{\mytag{n=2}{eq:taylor2}}-\frac{v_p^2}{3\tau_r}t^3}_{\mytag{n=3}{eq:taylor3}}+\frac{v_p^2}{12\tau_r^2}t^4}_{\mytag{n=4}{eq:taylor4}}+\mathcal{O}(t^5 ), 
\end{equation}

\noindent
we obtain different expressions depending on the desired degree of accuracy. In particular, we recover the already proposed expression for the second-degree approximation \ref{eq:taylor2} of previous paragraphs, but we could expand up to higher orders, theoretically infinite. Therefore, we might wonder what are the limits that we encounter when we apply this equation up to a specific term and what is the error associated to doing so.  \\

In order to check the accuracy of these equations for different examples, we simulated full experiments via an ABP model (see methods for more information) and we estimated the parameters from their MSDs according to Eq. \ref{eq:taylor}. These parameters were then compared to the actual values used in the simulations and the errors were computed. Each full experiment consisted of a simulation of 30 different active particles, moving with a certain constant speed $v_p$ and subjected to Brownian fluctuations depending on their size. Each trajectory was simulated for 30 s at 25 FPS, in order to mimic the typical conditions of an optical microscopy tracking experiment. This simple model does not consider small variations in the speeds of different particles or errors during the tracking of the particles. Therefore, the errors associated to the estimation of the parameters only come from over- or under-fitting of the equations by the different approximation, and are not experimental errors. In those cases, the final errors would be even higher than those reported here, which highlights the importance of choosing the right equation.\\

\begin{figure}[!t]
        \center{\includegraphics[width=0.95\textwidth]{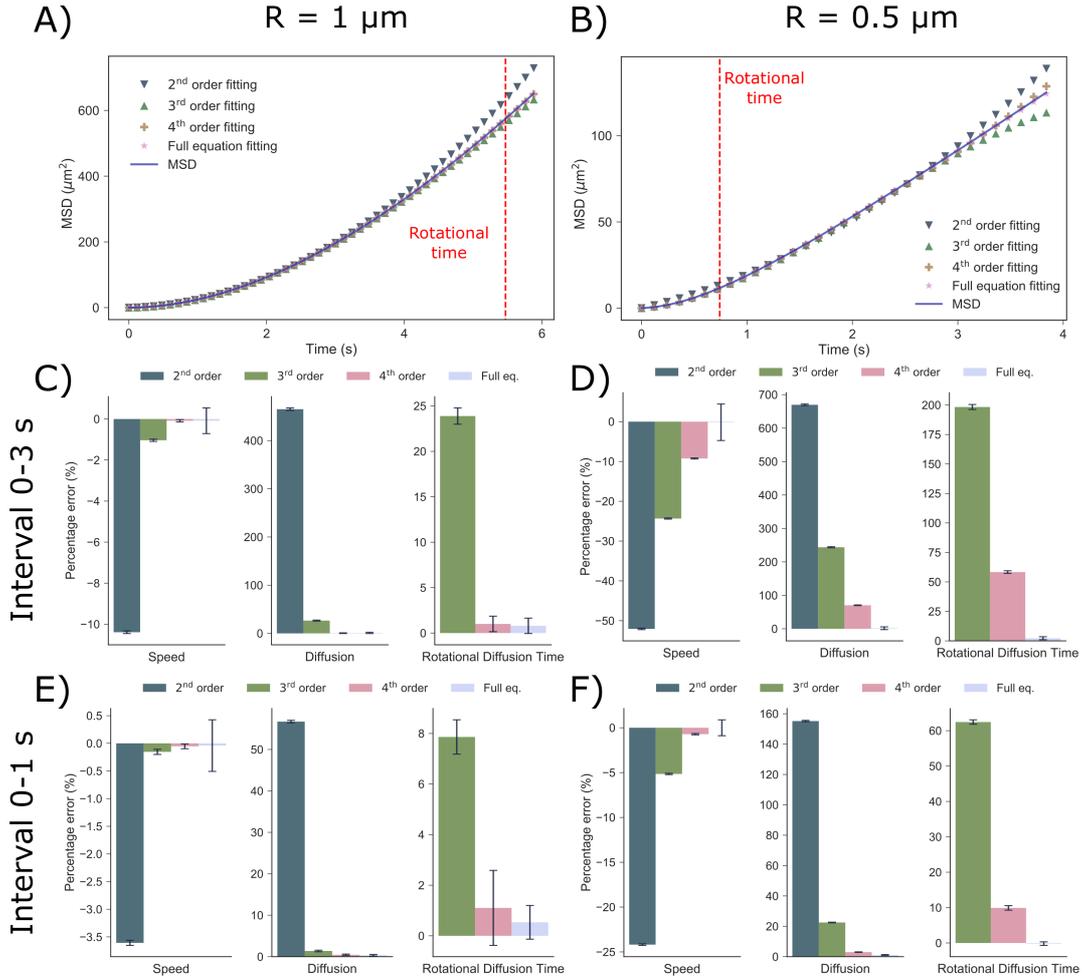}}
        \caption{\label{fig2} Errors associated to performing different polynomial approximations. A, B) Simulated MSD of a particle of $R = 1\,\mu$m and $R = 0.5\,\mu$m, respectively, moving at a speed of $5\,\mu $m/s, with different fittings. Red dashed lines the rotational diffusion time. C-F) Calculated error of the speed, diffusion coefficient and rotational diffusion time for different equations in the fitting intervals of [0,3] s and [0,1] s and for the two particle sizes above.}
\end{figure}

Generally, in the literature \cite{Hortelao2017a,Wang2019,Arque2019}, an interval between 1 s and 3 s is used to fit the MSD to Eq. \ref{eq:taylor}, \ref{eq:taylor2}, by using the least squares method. However, for the fitting to be precise, the upper bound of the fitting interval must be much smaller than the rotational diffusion time of the particles. An example of this is shown in Fig. \ref{fig2}A with an active particle of $R =1\,\mu$m moving at a speed of $5\, \mu $m/s. We found that a quadratic fitting underestimates the speed by 10\% and the diffusion coefficient by 500\%. This surprising result shows that the conditions for applying the second-degree approximation are not completely valid in cases like this one, where the fitting interval does not fulfill the requirement of Eq. \ref{eq:taylor}, \ref{eq:taylor2}. The rotational diffusion time, visible in Fig.  \ref{fig2}A, is too close to the fitting interval, and the condition $ t \ll \tau_r$ does not completely hold. If we perform a third- or fourth-degree polynomial fitting, we can reduce the error in the speed and diffusion coefficient to almost 0\%, while at the same time giving an estimation of the rotational diffusion time, which is not possible to obtain with a quadratic fitting. A fitting to the theoretical full equation can also yield correct results. However, this fitting is less recommended, as it shows greater error variability in the speed (it is not so stable) and it is extremely dependent on the initial parameters, which might require fine tuning before obtaining accurate values. \\

In the case of an active particle of $R = 0.5\, \mu $m moving at a speed of $5\, \mu $m/s, the associated errors of the fittings are even higher, due to the fact that the rotational diffusion time lies within the fitting interval. By performing a fourth degree fitting, the error of the speed can be reduced from 50\% (in quadratic fitting) to 10\%, and the diffusion coefficient error from roughly 650\% to less than 100\%. In this type of cases, the most recommended strategy to improve even more the estimation accuracy is to reduce the fitting interval, if there are enough data points to do so. Thus, the underestimation of the speed of a $R=1\, \mu$m particle can be reduced from 10\% (Fig.  \ref{fig2}C) to 3.5\% (Fig.  \ref{fig2}E) in the quadratic fitting. However, the diffusion coefficient is still overestimated to almost 60\% of its value. Again, a third- and fourth-degree fitting can reduce the errors to $\sim$0\% and allow an estimation of the rotational diffusion time. Similar results are obtained for the smaller particle in Fig.  \ref{fig2}F. \\

To summarize, choosing a first- or second-degree equation to fit the MSD is clearly context-dependent, and the choice depends on the size of the particle, its speed and the fitting interval. In the cases just shown, it can give over-estimations of  $D_t$ by 650\% or more, and under-estimations of the speed by 50\%. These errors could be larger for different scenarios, namely when the propulsive speed is much higher (Fig.  S1-4). However, we can conclude that, in general, reducing the fitting interval (if it is possible) can improve the accuracy of the estimations in every case. Moreover, we show that the polynomial fittings do not seem to be greatly affected by the number of particles of the total trajectory length (Fig. S5-8). Therefore, the error associated to the intrinsic variability of the MSD for small data sets seems to be smaller than the error coming from the polynomial approximations. In any case, it is always advisable to record as many particles as possible for as long as possible, since experimental errors, not accounted for here, can have a strong impact in the data. In general, third- and fourth-degree approximations provide a great improvement with respect to a quadratic approximation and can also give an estimation of the rotational diffusion time of the particle. Nevertheless, it needs to be emphasized that reporting the translational diffusion for micron-sized particles moving with high speeds will always be prone to error, since the Brownian fluctuations play a smaller role than the active motion. Presenting and comparing speeds of these particles will always be more advisable and, in order to report diffusion, the linear part of the MSD, long after the rotational diffusion time, should be fitted to a linear equation to extract the enhanced diffusion coefficient, following the equation $\mathrm{MSD}(t)=4D_e t$. Likewise, if the particles are very small (up to only a couple hundreds of nanometers in radius), their motion is mainly characterized by enhanced diffusion, and a simple linear fitting is the most appropriate approach, since none of the previous approximations can be applied. In the case of fitting to higher orders than \ref{eq:taylor2} of the Taylor expansion, we strongly suggest to use \ref{eq:taylor3}, since for each new order more terms need to be fitted and the result may strongly depend on the initial seed, as in the case of the full equation fitting.

\subsection{Active particles with exponentially decreasing speed}\label{sec:decay}

These approximations only hold in the ABP model, that assumes that the propulsive speed is constant. However, since catalytic micro- and nano-motors depend on their fuel to move, any change in the fuel concentration can lead to a change in their propulsion speed making this assumption incorrect.  For some types of active particles, such as light-powered or electric-field-dependent ones, this might not be an issue since the system is continuously fed, although it could occur when the motion source is removed or if the motion mechanism is complex \cite{Xuan2018}. For many catalytic reactions, where the fuel is added only at the beginning, fuel consumption or changes in the environment (e.g. pH) can make the speed change with time. This non-constant propulsive speed has been observed, for example, in enzymatically propelled micro-motors \cite{Patino2019}, and could also be observed in Pt-based motors. Therefore, understanding the parameters that govern this kind of dynamics can be helpful towards the comprehension of their motion mechanisms. Focusing on the example of enzymatically propelled micro-motors that lose speed exponentially, we solved Eq. \ref{eq:ABP} with a non-constant speed term following $v_p=v_0e^{-\beta t}$, where $v_0$ is the initial speed of the particle and $\beta$ is the ratio at which its speed decays.\\

In this case, obtaining the MSD is not straightforward because the system is no longer ergodic. Since the speed of the particle is not constant in time, the ensemble-average MSD and the time-average MSD will not be the same and strongly depend on the initial speed of the particle, $v_0$, as well as the total length of the trajectory (since time-average covers the whole trajectory, with a different speed at each point). The MSD of a single particle with an initial speed $v_0,$ recorded for a time $T$, would have the form (see Supplementary Information for more details):

\begin{equation}
\begin{aligned}
\textrm{TAMSD}( t) &= \frac{1}{T- t}\int_{0}^{T- t}dt_0\mathrm{MSD}(t_0, t)=\\  
&=\frac{v_0^2}{D_\theta^2-\beta^2}\bigg[2e^{- t(\beta+D_\theta)}-e^{-2\beta  t}-1+\\
&\frac{D_\theta}{\beta}\left(1-e^{-2\beta  t}\right)\bigg]\frac{1- e^{-2\beta(T-  t)}}{2\beta(T- t)}+ 4D_t  t,\\ \label{eq:ABP_v_solved} 
\end{aligned}
\end{equation}

\begin{figure}[!tb]
	\center{\includegraphics[width=0.92\textwidth]{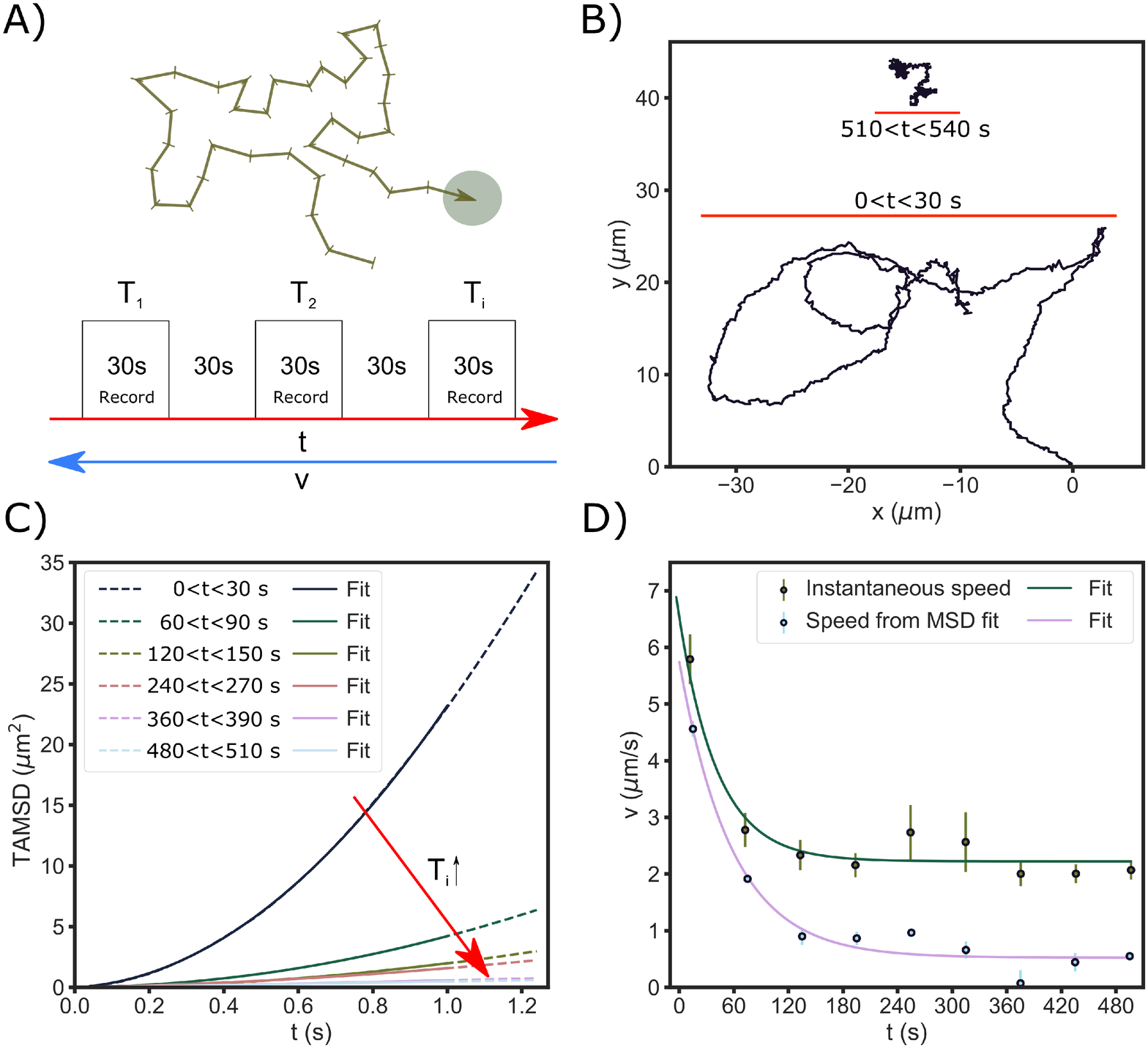}}
	\caption{\label{fig:decay} Self-propelled particles with exponentially decreasing speed. A) Schematic of the time splitting. We record 9 min videos and split it in videos of 30 s. During this time, we alternate 30 s of recording (referred as $T_i$) and 30 s of stopping. The later $T_i$, the less propulsive is the particle. B) Trajectory of a R = 1 $\mu$m particle recorded for 9 min. Here we present data for T1 ( [0,30) s) and T9 ( [480,510) s). Particle's speed decreases since for equal times, the particles travels a bigger distance in the beginning than in the end. C) MSD of the same particle in B, but for different intervals $T_i$ (dashed lines). These MSDs were fitted to a forth-term polynomial only up to 1 s. D) Speed of the same particle as in C). Two different speeds are presented: the propulsive speed taken from the MSD fit in C) and the average of the instantaneous speed for each $T_i$. In both cases, the speed decreases exponentially.}
\end{figure}

In practice, this equation would be too complex to fit with experimental data, resulting in a dead-end. However, here we propose an approximation of this equation, based on splitting the trajectory in smaller segments, that would allow us to extract the motion parameters, even if the particle loses speed fast. To show this, we present in Fig. \ref{fig:decay} an experimental example of an enzymatically propelled micro-particle of $R = 1\,\mu$m with a time-decreasing speed, based on a system from a previous work \cite{Patino2019}. In this case, we recorded a 9 min video and split it in a sequence of periods of 30 s where we alternated recording and pausing to re-focus the particle. We refer to each period of recording as T$_i$ as in Fig. \ref{fig:decay}A. In Fig. \ref{fig:decay}B we show the trajectory of this particle for the first 30 s and the last 30 s of recording, showing a clear decrease in speed after 500 s, since the trajectory has shrunk. A reduction in the covered area means that the MSD will also decrease over time, as represented by the dashed lines in Fig. \ref{fig:decay}C, which decrease in each recording step. By fitting each MSD (solid lines) and extracting the speed, we can see how it decays exponentially with time (Fig. \ref{fig:decay}D) until it reaches a near 0 propulsive speed. This effect is also observed after calculating the instantaneous speed of the particle along each video. We can see how the instantaneous speed also decreases exponentially with time, but it has a plateau different from 0, since the instantaneous speed also considers the effect that arises from Brownian motion.\\

In order to find a method of analysis that deals with this kind of time-dependent behaviour, we can try to expand Eq. \ref{eq:ABP_v_solved} with a Taylor approximation. In that case, we can separate it into three different components: i) a linear term coming from the Brownian fluctuations; ii) an exponential term (between square brackets) that strongly depends on the parameter $\beta$ and $D_{\theta}$; and iii) a correction term that depends on $\beta$ but also on the length of the video, $T$, termed $C(T, \beta,  t)$. Since $\beta$ is usually a small parameter (in our experiment $\beta\sim \mathcal{O}(-2)$ s$^{-1}$) and considering that $ t/\tau_r\sim0$, we can perform a Taylor expansion on the exponential terms between square brackets, from which we obtain, up to third order:

\begin{equation}
\textrm{MSD}( t) = 4D_t t + v_0^2 \bigg[ t^2 - \frac{1}{3}\left(D_R+3\beta\right)   t^3\bigg]C\left(T,\beta, t\right), \label{eq:expansionExp} 
\end{equation} 

\noindent
where we define the correction factor as:

\begin{equation}
C(T, \beta,  t) = \frac{1}{T- t}\left(\frac{1- e^{-2\beta(T- t)}}{2\beta}\right). \label{eq:correctionFactor} 
\end{equation}

From this approximation, several approaches can be followed depending on the characteristics of the system. If the parameter $\beta$, that characterizes how fast the particle loses speed, is very small compared to $D_R$, we recover the third degree approximation developed in Eq. \ref{eq:taylor}, \ref{eq:taylor3}. If the parameter $\beta$ is large, meaning that the particle loses speed very fast, the calculation of the MSD will not be very accurate, since its speed will be very different at the beginning and the end of the trajectory and the values of $r(\Delta t)$ would be incorrectly averaged. This undesirable effect could be corrected by recording short videos in which the speed does not change significantly and the MSD could be more reliably calculated, as we report in Fig. \ref{fig:decay}. However, we want to emphasize here that, even if we divide the whole trajectory into smaller segments, the system remains non-ergodic, and we are simply deciding to approximate it as a ergodic case. Ultimately, how much the MSD is affected by the exponentially decaying speed will depend on each specific case, and also on the length of the recorded trajectories, so the effect of the correction factor function should be carefully analyzed. \\

Therefore, it is important to understand that the deviation from the constant speed case can change the analysis of the MSD to obtain reliable results. In this particular case of the exponentially decaying speed, it is especially difficult to extract motion parameters in an accurate manner, not only to compare active particle's motion with different fuel concentrations, but also to understand their lifetime and motion mechanisms, and to gain feedback on how to perform the experiments. As a first approach to understand the dynamics of a new kind of active particle, the first recommendation would be to observe the long time behavior of them and extracting the speed profile from short-timed MSDs, as reported in Fig. \ref{fig:decay}. If the $\beta$ parameter is much smaller than $D_R$, then it is safe to assume that the particle does not lose speed fast enough to affect the results obtained from the MSD analysis. If $\beta$ cannot be ignored, as in our experimental examples, where $\beta\sim \mathcal{O}(-2)$ s$^{-1}$ , the correction factor (Eq. \ref{eq:correctionFactor}) plays a significant role on the motion analysis using TAMSD. In this case, plotting this function in terms of different trajectory lengths, $T$, can help us understand the possibilities and limitations of the analysis. \\

\begin{figure}
	\center{\includegraphics[width=0.98\textwidth]{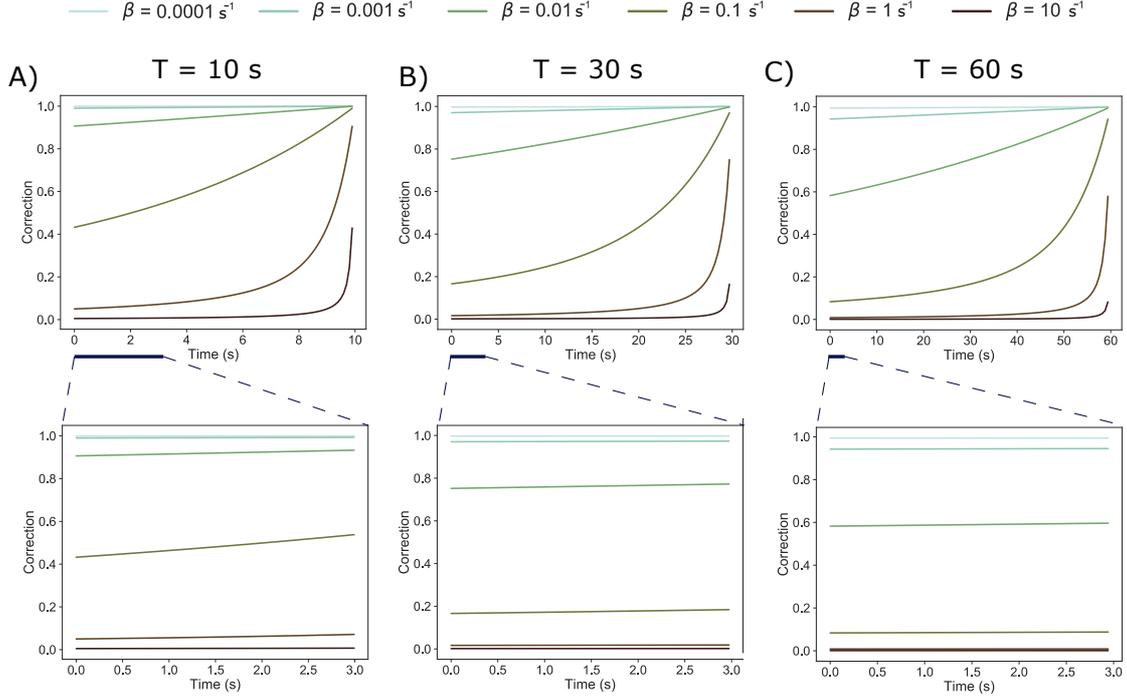}}
	\caption{Correction factor as a function of time, for three video lengths: A) 10 s; B) 30 s; and C) 60 s. Bottom plots show a close up of the first 3 s of the correction factor, which would apply to a typical MSD, showing how, in general, it can be considered a constant factor.}\label{fig:correction}
\end{figure}

For that reason, Fig. \ref{fig:correction} shows different scenarios of the correction term for several values of $\beta$ and trajectory lengths. We can see, for instance, that this function is in general not linear over its time domain, which makes it a complex issue to approximate or simplify the analysis and extraction of parameters. In particular, for a value $\beta = 0.01$  s$^{-1}$ (the order of magnitude of our experimental case), a correction factor of approximately 60\% would applied in the initial instants of the MSD for a video of 60 s, meaning that the magnitude of the MSD would be 60\% smaller than that of a particle moving constantly at the same speed. However, decreasing the length of the trajectory down to 10 s can increase this multiplicative factor up to 90\%. Nevertheless, although the correction factor is highly non-lineal,  only the initial part of the MSD has low variability (1/10th as a convention). Interestingly, this correction factor  remains constant during the first few seconds of the MSD, which is our region of interest (bottom close-ups of Fig. \ref{fig:correction}). This means that the corrected MSD in that time range could be approximated by:

\begin{equation}
\textrm{MSD}(\Delta t) = 4D_t\Delta t + v_0^2 \bigg[\Delta t^2 - \frac{1}{3}\left(D_R+3\beta\right)  \Delta t^3\bigg]C_{T,\beta} \phantom{........} \textrm{if } 0 < \Delta t \ll 3,  \label{eq:approximationExp} 
\end{equation} 

\noindent where $C_{T,\beta}$ represents the correction factor that is constant in time and is defined by the values of $T$ and $\beta$. This approximation and the values of the correction factor displayed in Fig. \ref{fig:correction} tell us that a video length as short as possible (or dividing a long one into smaller segments) ensures that the MSD remains "uncorrected" even for particles losing their speed fast. However, if it is too short, the correction factor might not be linear and the extraction of parameters would not be accurate, as for $T = 10$ s in Fig. \ref{fig:correction}A. For our experimental case of $\beta = 0.01$  s$^{-1}$ a video length of 30 s was shown to be adequate, as the correction factor remained around 0.75 for 3 s. Therefore, from Equation \ref{eq:approximationExp},  the actual speed extracted from the MSD would be underestimated by $\sqrt{0.75} = 0.87$. Indeed, we can see in the first data point of Fig. \ref{fig:decay}D that the velocity value extracted from the MSD is slightly underestimated compared to that obtained from the instantaenous speed (for large particles of $R = 1\ \mu$m with directional motion and high speed, the instantaneous speed can be considered representative; for lower speeds, as previosly mentioned, it is not reliable). It is important to bare in mind that, in any case, we are treating a non-ergodic system as an ergodic one, but we are compiling the non-ergodicity into that constant value that multiplies the non-linear factor of the MSD. Therefore, it is theoretically possible to use the MSD of a particle that loses speed to extract parameters, as long as the whole function is multiplied by this constant correction factor. However, it is still advisable to perform an ensemble average of several fresh particles, in order to decrease the variability and obtaining useful information about the system's dynamics. \\

\subsection{Active particles with angular speed}

Active particles intrinsically rotate due to its Brownian motion. This rotation is responsible for Brownian turnings in their trajectory and is characterized by $D_R$. Besides this Brownian rotation, some particles can also have active rotation, becoming active rotors. Active rotation is present when an active particle is chiral due to its body geometry \cite{Kummel2013}, non-uniformity in their fabrication \cite{Wang2017}, or the clustering of several active particles \cite{Ebbens2010}. Therefore, in many experimental cases, reporting the angular velocity of the particle can be useful, as this parameter is closely related to the propulsive speed. As an example, we performed experiments on enzymatically propelled micro-motors of $R = 1\,\mu$m that showed chiral motion, most likely due to anisotropies in their surface. In Fig. \ref{fig:rotational}A we present the trajectory of one of these particles. Notice how, due to its chirality, its trajectory shows circular patterns.\\

The study of the angular velocity of a particle requires to introduce new terms into Eq. \ref{eq:ABP}. As a first approach, since these kinds of particles lose speed we considered only the motion in short regimes of $[0,30]$ s. In these periods, we assume that the particle has a constant rotational speed and, hence, we need some extra constant in their equation of motion. To add this feature, we can change the rotational component in Eq. \ref{eq:ABP}, following:

\begin{figure}[!tb]
        \center{\includegraphics[width=\textwidth]{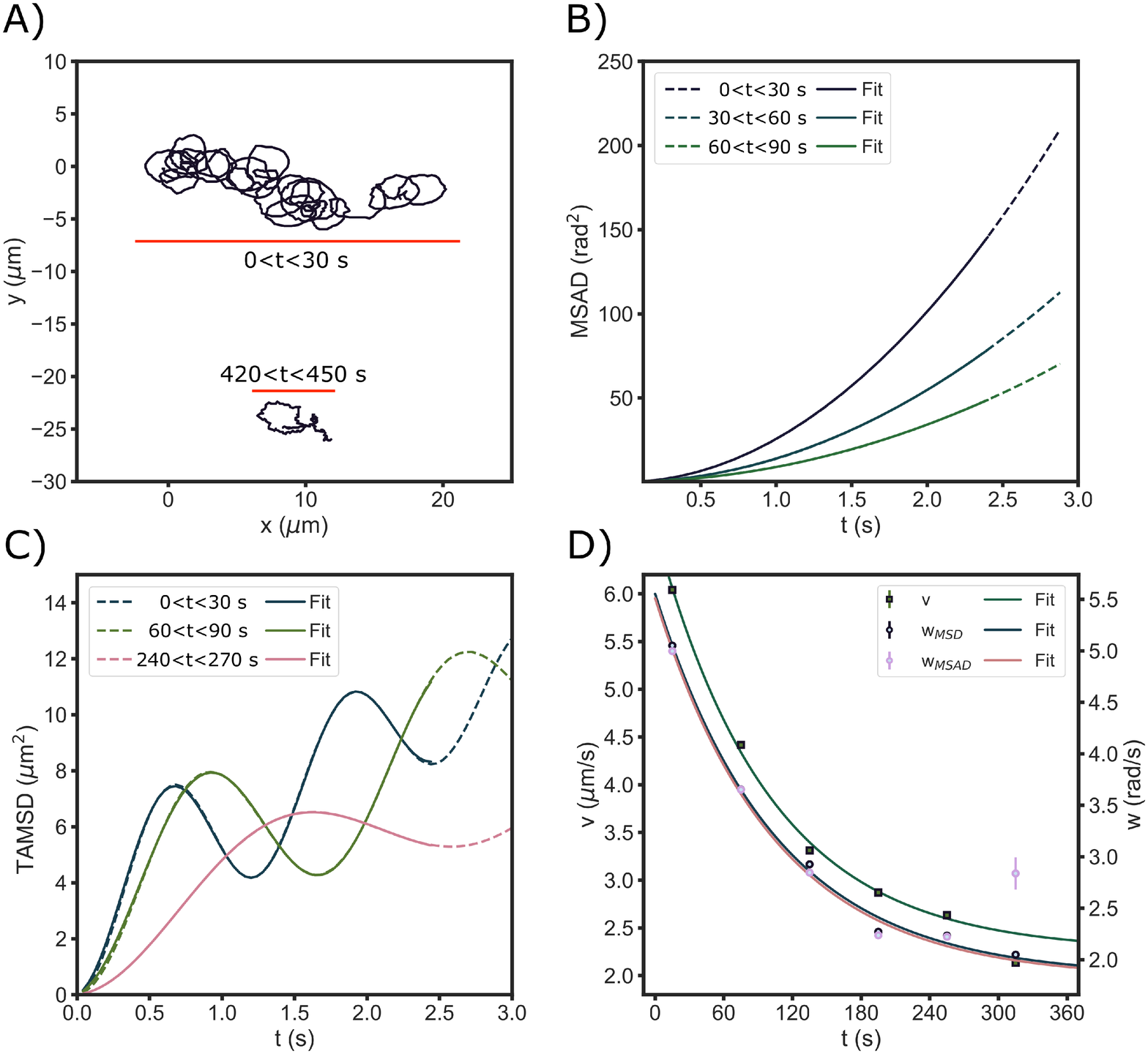}}
        \caption{\label{fig:rotational} Self-propelled particles with active rotation. A) Trajectory of a particle of R = 1 $\mu$m recorded for 450 s which exhibits active rotation. Here we considered the same schema for video splitting as in Fig. \ref{fig:decay}. Circular orbits are characteristics of this kind of activity. B) MSAD of the same particle as in A) for several time intervals. MSAD will be quadratic if particle is actively rotating with a constant speed. C) MSD of the particle presented in A) for different time intervals. Data is fit to the full Eq. \ref{eq:ABP_w_solved} D) By fitting the MSD of different time intervals to Eq. \ref{eq:ABP_w_solved} we can obtain the values for speed v and angular speed $\omega$. Here we can see how these values are decreasing exponentially in time. Moreover, we can see that $\omega$ calculated by MSD and MSAD are quite similar. In this case, last points for this particle may differ due to the bad data obtained for angular orientation.}
\end{figure}

\begin{equation}\label{eq:ABP_w}
	\dot{\theta}(t) = \omega  + \sqrt{2D_R}\xi_\theta(t),  
\end{equation}  

\noindent
where here, the new term $\omega$ is a constant angular speed. The MSD of this new equation is still ergodic and has already been discussed in the literature \cite{Ebbens2010}, yielding:  

\begin{equation}\label{eq:ABP_w_solved} 
\begin{split} 
	\mathrm{MSD}( t) = 4D_tt+\frac{v_0^2}{D_R^2+\omega^2}\bigg( 2D_Rt + &\frac{2}{D_R^2+\omega^2}\bigg[e^{-D_Rt}\cos(\omega t)(D_R^2-\omega^2) \\ 
 &-2\omega D_Re^{-D_Rt}\sin(\omega t) - D_R^2 + \omega^2 \bigg]\bigg).
\end{split} 
\end{equation}

However, further simplifications or approximations of these equations, as those discussed for non-chiral particles, have not been considered. In this scenario, the common time limits that worked for Eq. \ref{eq:taylor}, based on $\tau_r$, do not give useful approximations to fit to experimental data. To better fit this equation, we need to make  additional approximations. From Eq. \ref{eq:ABP_w_solved}, first we can notice  that the Brownian rotational diffusion is accompanied by the addition or subtraction of the active angular rotation, $\omega$. We can therefore think of an extra limit where $\omega \gg D_R$ or $\omega \ll D_R$ in order to simplify this equation. To know if we are in one of these scenarios, we recommend visualizing the mean squared angular displacement (MSAD) when possible, as it can provide interesting insights into the rotational dynamics of the particles \cite{Das2015,Simmchen2017}. The solution of the MSAD for Eq. \ref{eq:ABP_w} is an easy quadratic formula:  

\begin{equation}\label{eq:ABP_w_MSAD}  
	\mathrm{MSAD}( t) = 2D_Rt + \omega^2 t^2,  
\end{equation}  

\noindent
for which both constants can be fitted. To calculate this MSAD it is necessary to define the orientation angle $\theta$ of the particle, which is easy to calculate if the particle has some visual anisotropy. Otherwise, the direction can be defined from the velocity vector of the particle, calculated from the instantaneous velocity. Once $\theta$ is computed, the calculation of MSAD is analogous to the MSD, using $\theta$ instead of the position. Notice that in MSAD we do not use confined angles in [0:2$\pi$] rad, but cumulative. In Table \ref{tbl:ABP_w} we present some approximations depending on the ratios $D_R/t$ and $\omega/D_R$. We can see how for long times compared to the rotational diffusion time $\tau_R$, the motion becomes diffusive with a linear MSD  with enhanced diffusion, even if the nano- or micro-motor shows very high chirality. Here, we can distinguish two cases: if $D_R/\omega \ll 1$, the enhanced diffusion is inversely proportional to the value of $\omega$. If $\omega$ is very large, $D_e \approx D_t$, that is, the rotation of the particle will prevent it covering a wide area, despite its activity. If $D_R/\omega \gg 1$, there is no such dependency on $\omega$. In the regime where we would find a ballistic term for Eq. \ref{eq:ABP} ($t\ll\tau_R$), we can see how the expression has several cosines functions that represent the rotations of the particle. Therefore, finding an MSD with oscillations would give indications that the active particles show active rotation, as shown in Fig. \ref{fig:rotational}C. In this case, if the condition $t\ll\tau_R$ is fulfilled, the MSD could be fitted to this equation, simpler than Eq. \ref{eq:ABP_w_MSAD} to extract the relevant parameters of motion. Using a previous fitting to the MSAD equation to obtain an estimation of $\omega$ could be helpful to insert initial parameters for the MSD fitting. \\

\begin{table}[tb]   
	\centering   
	$\begin{array}{cll}   
	\toprule \mathrm{Ratio} & \multicolumn{1}{c}{\hspace{1cm}\mathbf{t \ll \tau_r}} & \multicolumn{1}{c}{\hspace{1cm}\mathbf{t \gg \tau_r}} \\  \hline  
	\midrule \addlinespace   
	\mathbf{D_R/\omega \ll 1} & \multicolumn{1}{c}{\hspace{1cm}4D_t t+\frac{2v^2}{\omega^2}\left[ D_Rt - \left(1-D_Rt+\frac{(D_Rt)^2}{2}\right)\cos(\omega t)+1\right]} &   
	\multicolumn{1}{c}{\hspace{1cm}4\left(D_t + \frac{v^2D_R}{2\omega^2}\right)t} \\   
	\addlinespace \midrule \addlinespace   
\mathbf{D_R/\omega \gg 1} & \multicolumn{1}{c}{\hspace{1cm}4D_t t+\frac{2v^2}{D_R^2}\left[ D_Rt + \left(1-D_Rt+\frac{(D_Rt)^2}{2}\right)\cos(\omega t)-1\right]} & \multicolumn{1}{c}{\hspace{1cm}4\left(D_t + \frac{v^2}{2D_R}\right)t} \\ \addlinespace \bottomrule \end{array}$    
	\caption{MSD approximations for chiral active particles and different ratios of $D_R/t$ and $\omega/D_R$.}    
	\label{tbl:ABP_w} 
\end{table}   

In Fig. \ref{fig:rotational}B and Fig. \ref{fig:rotational}C we present an example of the calculated MSAD for the particle presented in Fig. \ref{fig:rotational}A calculated for different time regions of the video and how Eq. \ref{eq:ABP_w_MSAD} fits well with the experimental data. Although our initial assumption of constant angular rotation can be true for some kind of catalytic particles \cite{Ebbens2010}, in our experimental case with an enzymatic particle this is not correct. We can see how for every time interval, the trajectory is different (see Fig. \ref{fig:rotational}A) due to a decrease of both propulsive and angular speed. In order to include this behavior together with chirality, we should mix the previous results shown for particles with decaying speed and obtain new equations. Since these results were quite complex, we discarded the calculus of these combined formulas. Instead, we decided to perform analysis separately for each time interval and hypothesize that both the angular and the propulsion speed are constant in each 30 s regime. 

 Since the rotation and speed are slowing down, the MSD will show that, at larger times, the fluctuations have larger amplitude and the value of the MSD is generally lower. In Fig. \ref{fig:rotational}D one can see that both the linear  and  angular speed decay exponentially with time. In particular, we can see that there is a coupling between the angular and the linear speed -- if the propulsive speed decreases, the active rotation also decreases following the same trend. Data calculated from MSD must agree with data calculated from MSAD, as seen in Fig. \ref{fig:rotational}D  for $\omega$. For the last values we can see a mismatch between MSD and MSAD. This mismatch is reasonable since in our case, we calculate our angle from the velocity vector. If the speed is very low, the angle taken from this method will show considerable error, which means that MSAD will be affected and, hence, also $\omega$.  Finally, we would also like to indicate that even in a case where propulsive and rotational speeds are constant in time, fitting directly to Eq. \ref{eq:ABP_w_solved} would be a bad idea because of the complexity of this formula, as the results will strongly depend on the initial seed given to the software. For this reason, using the Equations from Table \ref{tbl:ABP_w}  can provide better approximations.

\subsection{Active particles with drift}

One of the main problems regarding motion analysis of active particles is  how to distinguish active motion from advection or drift \cite{Dunderdale2012a,Byun2017} to be sure that the reported speeds actually come from the self-propulsion of the particle. For this reason, it should be checked if the overall population of particles is moving in the same direction or in random directions as reflected in Fig \ref{fig:drift}A. The first case would indicate that there is drift in the system, while the former will be a sign that there is not. In theory, it is possible to subtract the component of advection with post-processing techniques if many particles are recorded (and especially if there are passive tracers), and the data could still be useful. However, if only one or few of them were recorded, the presence of drift would mask the effects of active motion. Although checking the overall population of particles should be the first action to take, it is always recommended to check for drift during analysis by comparing the MSD with a new modification of Eq. \ref{eq:ABP}. To introduce the effect of drifting, we consider that drift induces a constant force over particles in a specific constant direction as: 

\begin{figure}[!tb]
        \center{\includegraphics[width=\textwidth]{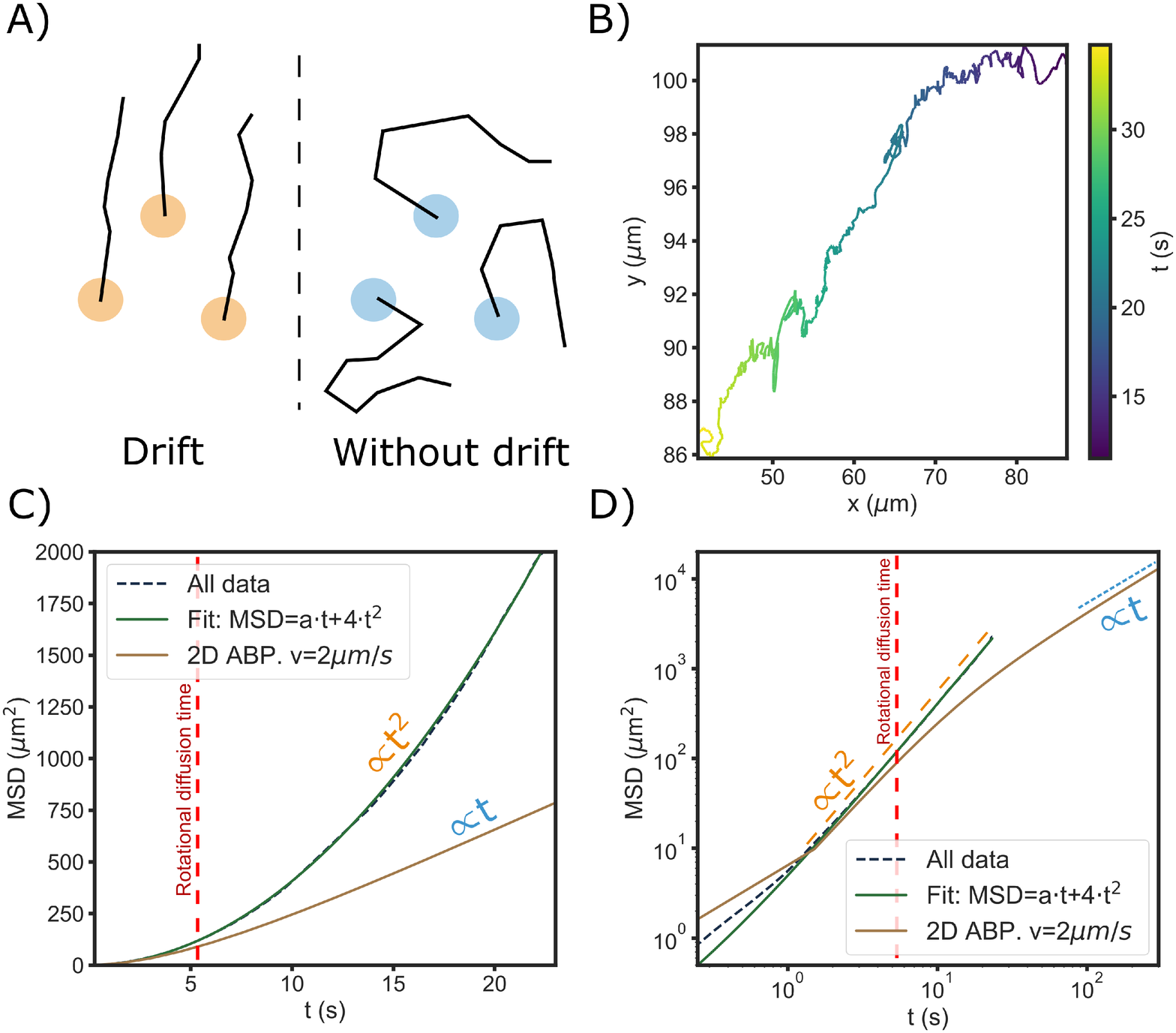}}
        \caption{\label{fig:drift} Particles with drift. A) Particles that experiment drift move all in the same direction (left) while particles not affected by drift experiment movement in different directions (right). B) Trajectory of a recorded passive particle of R = 1 $\mu$m undergoing drift. C) MSD plot of the particle recorded and presented in B (dashed and green lines) and the theoretical equation of an active particle (brown) following Eq. \ref{eq:ABP} with the same speed as calculated with the drift case. D) Log-log of C) plot. The log-log view can help us also distinguishing scenarios between propulsive and Brownian easier. }
\end{figure}

\begin{equation}\label{eq:ABP_drift} 
\begin{aligned}
\dot{x}(t) &= v_p\cos(\theta(t)) + \sqrt{2D_t}\xi_x(t) + v_d\cos(\theta_d), \\ 
\dot{y}(t) &= v_p\sin(\theta(t)) + \sqrt{2D_t}\xi_y(t) + v_d\sin(\theta_d), \\ 
\dot{\theta}(t) &= \sqrt{2D_R}\xi_\theta(t), \\ 
\end{aligned}
\end{equation}

\noindent
where here $v_d$ refers to the speed of the particle due to drift and $\theta_d$ is its drifting angle. Because of the drift, the new formula for the MSD is the same as in Eq. \ref{eq:MSDactive} but with the addition of an extra quadratic term: 

\begin{equation}\label{eq:ABP_drift_solved} 
	\mathrm{MSD}( t) = 4D_tt + 2\frac{v^2}{D_R^2}\left(D_Rt+e^{-D_Rt}-1\right) + v_d^2t^2 = \mathrm{MSD}_{\mathrm{No\ drift}} + v_d^2t^2. \\ 
\end{equation}

\begin{figure}[!tb]
        \center{\includegraphics[width=0.95\textwidth]{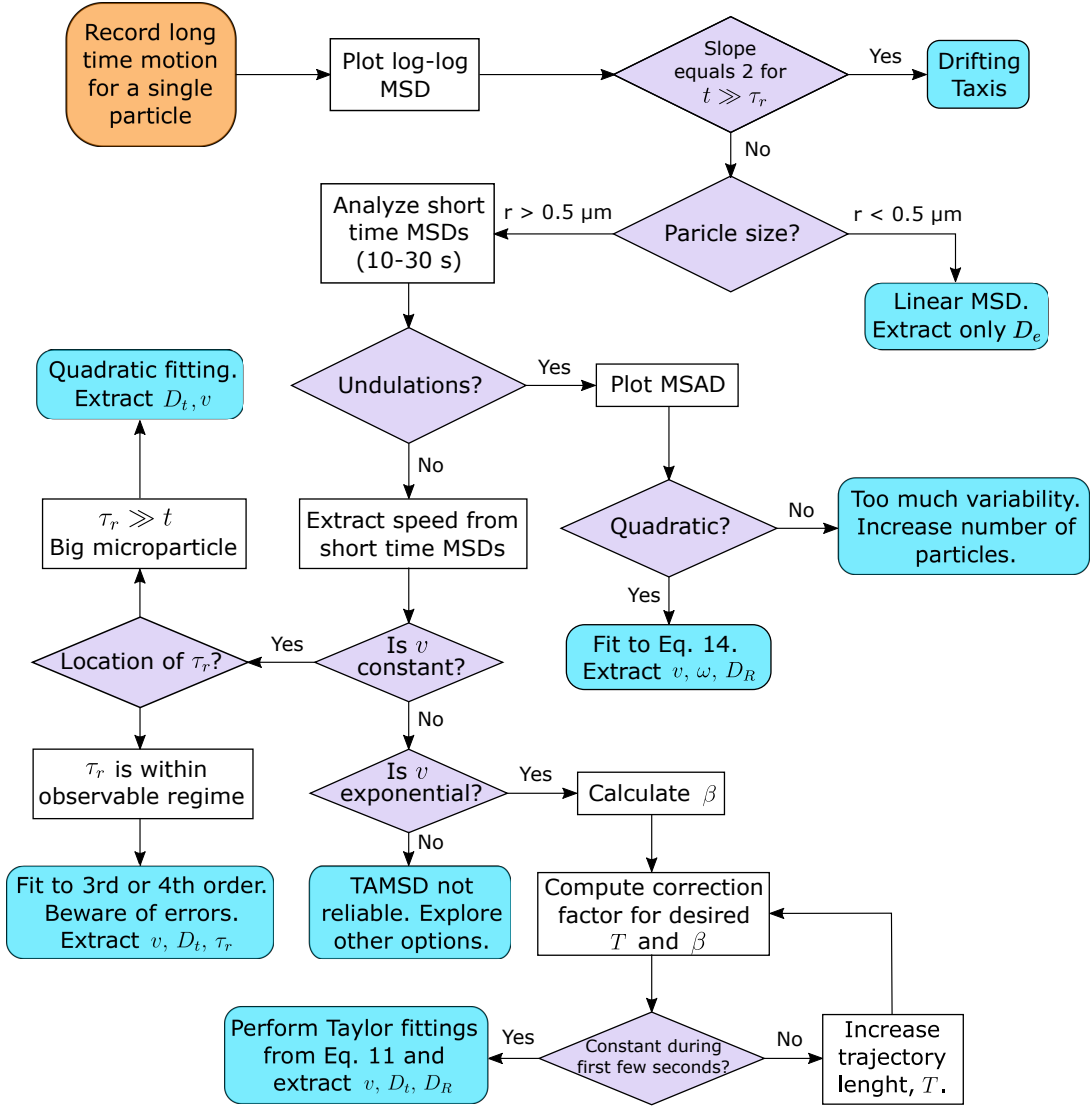}}
        \caption{\label{fig:scheme} Flowchart considering the different possibilities of analysis considered in this work and the parameters that can be extracted in each case.}
\end{figure}

The extra quadratic term that appears in Eq. \ref{eq:ABP_drift_solved} implies that, unlike in the scenario without drifting, at times longer than the rotational diffusion time, the MSD will look quadratic instead of linear with an enhanced diffusion. As an example, we present in Fig. \ref{fig:drift} experimental data of one inactive silica micro-particle of $R= 1\,\mu$m where drift was induced. In Fig. \ref{fig:drift}B we can see how the direction of this particle is always the same. When the MSD is plotted (see Fig. \ref{fig:drift}C), its curve fits very well with a parabolic function even at large times compared with its rotational diffusion time. However, comparing it with a simulated active particle of the same size and $v=v_d$, we can see how after the rotational diffusion time, the MSD of the simulated active particle stops being quadratic and grows linearly, as the limits of Eq. \ref{eq:MSDactive} indicate. Following a quadratic tendency after the rotational diffusion time can be a clear indication of a drifting-like behavior in an active particle system.\\

In order to check the effects of drifting or tactic effects produced by a gradient in the motion of micro- and nano-motors, the long-term behavior of the MSD should be analyzed as indicated in the previous paragraph. An estimation of the rotational diffusion time, $\tau_r$, should already be known, either theoretically from the Stokes-Einstein equation, from a 3rd or 4th order fitting to the MSD  or extracted from the velocity autocorrelation. Then, a trajectory much longer than this time should be obtained and the MSD calculated. By plotting the MSD in a double log-log scale, we should see a change in slope at the rotational diffusion time, as in Fig. \ref{fig:drift}D. In this figure, we can see how the simulated active particle follows a quadratic tendency during the first seconds, before reaching the rotational diffusion time, and its slope decreases to 1 after a longer time. In general, doing a log-log plot is a good approach to evaluate the presence of drift in the system, as well as to investigate different motion from the MSD of micro- or nano-motors that move with sub- or super-diffusive dynamics not collected in any of our previous examples \cite{Patino2018a,Xuan2018}. The evaluated trajectory should be long enough so that the variability of the MSD can be kept to a minimum (that is, at least 10 times longer than the rotational diffusion time, as discussed in Section 4.1). If is known that no gradients or tactic effects are present in the system and the analysis displays quadratic behavior for times longer than $\tau_r$, drifting or undesired convention flows are the most plausible reason behind it. Moreover, when a particle is showing drift-like behavior, the error of the MSD should remain small, due to the high directionality of the motion, which is deterministic, compared to when it is being dominated by Brownian fluctuations. In that case, the ``full MSD" should be smooth and highly parabolic at long times, meanwhile the one of an active particle without drift would show non-uniform or sharp variations in its shape, as those in Fig. \ref{fig1}.

\section{Conclusion}

In this work, we revise the current analysis methods for speed detection of self-propelled micro- and nano-motors based on optical tracking techniques, a crucial parameter to characterize their motion, and we provide new tools and approaches to minimize the errors and improve the quality of the results, taking into account different scenarios, which can be summarized with the flow chart shown in Fig \ref{fig:scheme}. Initially, we emphasize  that the calculation of the instantaneous speed generally used to compare different types of motion  is unreliable for small catalytic motors and depends on the parameters of the microscope. Moreover, we analyze other sources of error originated from the classical approximations of the MSD equation of an active Brownian particle, which is considered the most robust method to extract the propulsive speed or the translational diffusion of the particle. We find that, although the error of the estimation of the speed is usually not very high, the translational diffusion coefficient can be extremely overestimated by 600\% or more, especially for particles of medium size (between $R = 0.5-1\,\mu$m), where the rotational diffusion time is within the observable regime. We propose an alternative analysis based on a simple Taylor expansion of higher orders, which can reduce the error of the parameters theoretically to almost 0\% (leaving only the systematic error associated with the recording and tracking process) and allows to even estimate the rotational diffusion time, therefore increasing the amount and quality of the information to be extracted. \\

Finally, we focus on three experimental scenarios of catalytic motors with different types of velocity profiles and study how their analysis should be approached. First, we consider the case of ergodicity breaking caused by an exponentially decaying speed, for which we propose several approaches and approximations to fully understand and characterize this kind of system in an accurate manner. We also consider the case of an active particle with active angular speed, together with propulsive speed, and we study the limiting cases of this scenario and propose new equations, depending on the values of $D_R$ and $\omega$, and how they affect the MSD. Lastly, we tackle the example of a particle under drifting or tactic behavior under substrate gradients and how they can be distinguished from classical active Brownian motion without directionality. We strongly believe that these results will provide useful tools to improve the quality of the analysis of micro- and nano-motors powered by catalytic reactions and help in the in the study of their sources of motion by an improved extraction of motion parameters. 

\section{Acknowledgements}
The research leading to these results has received funding from the Spanish MINECO for grants CTQ2015-68879-R (MICRODIA) and CTQ2015-72471-EXP (Enzwim). S.S. acknowledges Foundation BBVA for the MEDIROBOTS project, the CERCA program by the Generalitat de Catalunya and the project RTI2018-098164-B-I00 financed by Ministerio de Ciencia, Innovación y Universidades (MCIU), Agencia Estatal de Investigación (AEI) and Fondo Europeo de Desarrollo Regional (FEDER). I. P. acknowledges funding from the European Union’s Horizon 2020 program under ETN Grant Agreement No. 674979-NANOTRANS. R.M. thanks “la Caixa” Foundation through IBEC International PhD Programme “la Caixa” Severo Ochoa fellowships (code LCF/BQ/SO16/52270018). L.P. thanks MINECO for the FPI BES-2016-077705 fellowship. X.A. thanks MINECO for the Severo Ochoa programme (SEV-2014-0425) for the PhD fellowship (PRE2018-083712). S.S. and I.P. supervised the work. R.M. and L.P. performed the theory, simulations and wrote the paper. X.A. performed experiments. A.M-L. contributed with theory and discussions. All authors have discussed and approved the manuscript.

\bibliography{MSDpaper}

\end{document}